\documentstyle[twoside]{article}

\catcode`\@=11
\long\def\@makefntext#1{
\protect\noindent \hbox to 3.2pt {\hskip-.9pt
$^{{\eightrm\@thefnmark}}$\hfil}#1\hfill}               

\def\@makefnmark{\hbox to 0pt{$^{\@thefnmark}$\hss}}    

\def\ps@myheadings{\let\@mkboth\@gobbletwo
\def\@oddhead{\hbox{}
\rightmark\hfil\eightrm\thepage}
\def\@oddfoot{}\def\@evenhead{\eightrm\thepage\hfil
\leftmark\hbox{}}\def\@evenfoot{}
\def\sectionmark##1{}\def\subsectionmark##1{}}

\oddsidemargin=\evensidemargin
\newcounter{sectionc}\newcounter{subsectionc}\newcounter{subsubsectionc}
\renewcommand{\section}[1] {\vspace{12pt}\addtocounter{sectionc}{1}
\setcounter{subsectionc}{0}\setcounter{subsubsectionc}{0}\noindent
        {\tenbf\thesectionc. #1}\par\vspace{5pt}}
\renewcommand{\subsection}[1] {\vspace{12pt}\addtocounter{subsectionc}{1}
        \setcounter{subsubsectionc}{0}\noindent
        {\bf\thesectionc.\thesubsectionc. {\kern1pt \bfit #1}}\par\vspace{5pt}}
\renewcommand{\subsubsection}[1] {\vspace{12pt}\addtocounter{subsubsectionc}{1}
        \noindent{\tenrm\thesectionc.\thesubsectionc.\thesubsubsectionc.
        {\kern1pt \tenit #1}}\par\vspace{5pt}}
\newcommand{\nonumsection}[1] {\vspace{12pt}\noindent{\tenbf #1}
        \par\vspace{5pt}}

\newcounter{appendixc}
\newcounter{subappendixc}[appendixc]
\newcounter{subsubappendixc}[subappendixc]
\renewcommand{\thesubappendixc}{\Alph{appendixc}.\arabic{subappendixc}}
\renewcommand{\thesubsubappendixc}
        {\Alph{appendixc}.\arabic{subappendixc}.\arabic{subsubappendixc}}

\renewcommand{\appendix}[1] {\vspace{12pt}
        \refstepcounter{appendixc}
        \setcounter{figure}{0}
        \setcounter{table}{0}
        \setcounter{lemma}{0}
        \setcounter{theorem}{0}
        \setcounter{corollary}{0}
        \setcounter{definition}{0}
        \setcounter{equation}{0}
        \renewcommand{\thefigure}{\Alph{appendixc}.\arabic{figure}}
        \renewcommand{\thetable}{\Alph{appendixc}.\arabic{table}}
        \renewcommand{\theappendixc}{\Alph{appendixc}}
        \renewcommand{\thelemma}{\Alph{appendixc}.\arabic{lemma}}
        \renewcommand{\thetheorem}{\Alph{appendixc}.\arabic{theorem}}
        \renewcommand{\thedefinition}{\Alph{appendixc}.\arabic{definition}}
        \renewcommand{\thecorollary}{\Alph{appendixc}.\arabic{corollary}}
        \renewcommand{\theequation}{\Alph{appendixc}.\arabic{equation}}
        \noindent{\tenbf Appendix \theappendixc #1}\par\vspace{5pt}}
\newcommand{\subappendix}[1] {\vspace{12pt}
        \refstepcounter{subappendixc}
        \noindent{\bf Appendix \thesubappendixc. {\kern1pt \bfit #1}}
        \par\vspace{5pt}}
\newcommand{\subsubappendix}[1] {\vspace{12pt}
        \refstepcounter{subsubappendixc}
        \noindent{\rm Appendix \thesubsubappendixc. {\kern1pt \tenit #1}}
        \par\vspace{5pt}}

\newcommand{\smalllineskip}{\baselineskip=10pt}

\def\eightcirc{
\begin{picture}(0,0)
\put(4.4,1.8){\circle{6.5}}
\end{picture}}
\def\eightcopyright{\eightcirc\kern2.7pt\hbox{\eightrm c}}

\def\abstracts#1#2#3{{
        \centering{\begin{minipage}{4.5in}\baselineskip=10pt\footnotesize
        \parindent=0pt #1\par
        \parindent=15pt #2\par
        \parindent=15pt #3
        \end{minipage}}\par}}

\newcommand{\bibit}{\nineit}
\newcommand{\bibbf}{\ninebf}
\renewenvironment{thebibliography}[1]
        {\frenchspacing
         \ninerm\baselineskip=11pt
         \begin{list}{\arabic{enumi}.}
        {\usecounter{enumi}\setlength{\parsep}{0pt}
         \setlength{\leftmargin 12.7pt}{\rightmargin 0pt} 
         \setlength{\itemsep}{0pt} \settowidth
        {\labelwidth}{#1.}\sloppy}}{\end{list}}

\newcounter{itemlistc}
\newcounter{romanlistc}
\newcounter{alphlistc}
\newcounter{arabiclistc}

\newcommand{\fcaption}[1]{
        \refstepcounter{figure}
        \setbox\@tempboxa = \hbox{\footnotesize Fig.~\thefigure. #1}
        \ifdim \wd\@tempboxa > 5in
           {\begin{center}
        \parbox{5in}{\footnotesize\smalllineskip Fig.~\thefigure. #1}
            \end{center}}
        \else
             {\begin{center}
             {\footnotesize Fig.~\thefigure. #1}
              \end{center}}
        \fi}

\newcommand{\tcaption}[1]{
        \refstepcounter{table}
        \setbox\@tempboxa = \hbox{\footnotesize Table~\thetable. #1}
        \ifdim \wd\@tempboxa > 5in
           {\begin{center}
        \parbox{5in}{\footnotesize\smalllineskip Table~\thetable. #1}
            \end{center}}
        \else
             {\begin{center}
             {\footnotesize Table~\thetable. #1}
              \end{center}}
        \fi}

\def\@citex[#1]#2{\if@filesw\immediate\write\@auxout
        {\string\citation{#2}}\fi
\def\@citea{}\@cite{\@for\@citeb:=#2\do
        {\@citea\def\@citea{,}\@ifundefined
        {b@\@citeb}{{\bf ?}\@warning
        {Citation `\@citeb' on page \thepage \space undefined}}
        {\csname b@\@citeb\endcsname}}}{#1}}

\newif\if@cghi
\def\cite{\@cghitrue\@ifnextchar [{\@tempswatrue
        \@citex}{\@tempswafalse\@citex[]}}
\def\citelow{\@cghifalse\@ifnextchar [{\@tempswatrue
        \@citex}{\@tempswafalse\@citex[]}}
\def\@cite#1#2{{$\null^{#1}$\if@tempswa\typeout
        {IJCGA warning: optional citation argument
        ignored: `#2'} \fi}}

\def\@refcitex[#1]#2{\if@filesw\immediate\write\@auxout
        {\string\citation{#2}}\fi
\def\@citea{}\@refcite{\@for\@citeb:=#2\do
        {\@citea\def\@citea{, }\@ifundefined
        {b@\@citeb}{{\bf ?}\@warning
        {Citation `\@citeb' on page \thepage \space undefined}}
        \hbox{\csname b@\@citeb\endcsname}}}{#1}}

\def\@refcite#1#2{{#1\if@tempswa\typeout
        {IJCGA warning: optional citation argument
        ignored: `#2'} \fi}}

\def\refcite{\@ifnextchar[{\@tempswatrue
        \@refcitex}{\@tempswafalse\@refcitex[]}}

\def\pmb#1{\setbox0=\hbox{#1}
        \kern-.025em\copy0\kern-\wd0
        \kern.05em\copy0\kern-\wd0
        \kern-.025em\raise.0433em\box0}

\def\fnm#1{$^{\mbox{\scriptsize #1}}$}
\def\fnt#1#2{\footnotetext{\kern-.3em
        {$^{\mbox{\scriptsize #1}}$}{#2}}}

\def\fpage#1{\begingroup
\voffset=.3in
\thispagestyle{empty}\begin{table}[b]\centerline{\footnotesize #1}
        \end{table}\endgroup}

\def\runninghead#1#2{\pagestyle{myheadings}
\markboth{{\protect\footnotesize\it{\quad #1}}\hfill}
{\hfill{\protect\footnotesize\it{#2\quad}}}}
\font\tenrm=cmr10
\font\tenit=cmti10
\font\tenbf=cmbx10
\font\bfit=cmbxti10 at 10pt
\font\ninerm=cmr9
\font\nineit=cmti9
\font\ninebf=cmbx9
\font\eightrm=cmr8

\def\qed{\hbox{${\vcenter{\vbox{                        
   \hrule height 0.4pt\hbox{\vrule width 0.4pt height 6pt
   \kern5pt\vrule width 0.4pt}\hrule height 0.4pt}}}$}}

\begin{document}

\newcommand{\Vi}{\left(\sum\limits_i{\bf V}_i\cdot\nabla\right)}

\runninghead{Andrew E. Chubykalo}
{On the
Physical Origin of the Oppenheimer-Ahluwalia Zero-Energy Solutions $\ldots$}

\thispagestyle{empty}\setcounter{page}{1}
\vspace*{0.88truein}
\fpage{1}

\centerline{\bf ON THE PHYSICAL ORIGIN OF THE OPPENHEIMER-AHLUWALIA}
\vspace*{0.035truein}
\centerline{\bf ZERO-ENERGY SOLUTIONS OF MAXWELL EQUATIONS}

\vspace*{0.37truein}
\centerline{\footnotesize ANDREW E. CHUBYKALO}

\centerline{\footnotesize \it
Escuela de F\'{\i}sica, Universidad Aut\'onoma de Zacatecas}
\baselineskip=10pt
\centerline{\footnotesize \it
Apartado Postal C-580\, Zacatecas 98068, ZAC., M\'exico}


\baselineskip 5mm

\vspace*{0.21truein}

\abstracts{
In virtue of the Chubykalo - Smirnov-Rueda generalized form of
the Maxwell-Lorentz equation  a new form of the energy
density of the electromagnetic field  was obtained. This result allows us
to explain a physical origin of the Oppenheimer-Ahluwalia zero-energy
solutions of the Maxwell equations.}{}{}


\bigskip

$$$$
\section{Introduction}
\noindent
If $\phi_L({\bf p})$
and $\phi_R({\bf p})$ represent the massless $(1,0)$ and $(0,1)$
fields respectively [\refcite{BWW}], then the source-free
momentum-space
Maxwell equation can be written as
(see, e.g., Ref. [\refcite{dva_thesis}])\fnm{a}\fnt{a}
{The configuration-space Maxwell equations follow on setting
\[
{\bf p}=-i\nabla,\quad p^0=i\frac{\partial}{\partial t},
\]
and making appropriate linear combinations of the $\phi_R(x)$ and
$\phi_L(x)$ to obtain the $\bf E$ and $\bf H$ fields.
}
\begin{eqnarray}
\left({\bf J}\cdot{\bf p} +  p^0\right)\phi_L({\bf p}) &=&0 \label{a}\\
\left({\bf J}\cdot{\bf p} -  p^0\right)\phi_R({\bf p}) &=&0 \label{b}
\end{eqnarray}
where $\bf J$ are the $3\times 3$ spin-1 angular momentum matrices
\begin{equation}
J_x= \left(\begin{array}{ccc}
0 & 0 & 0 \\
0 & 0 & -i \\
0 & i & 0
\end{array}
\right),\quad
J_y=\left(\begin{array}{ccc}
0 & 0 & i\\
0 & 0 & 0\\
-i & 0 & 0
\end{array}
\right),\quad
J_z=\left(\begin{array}{ccc}
0 & -i & 0 \\
i & 0 & 0 \\
0 & 0 & 0
\end{array}
\right).
\end{equation}
Oppenheimer  [\refcite{Opp}] and Ahluwalia
[\refcite{dva_thesis},\refcite{dva_mpla},\refcite{ref3}] independently
noted that in order that non-trivial solutions of
Eqs. (\ref{a}) and (\ref{b}) exist one must have
\begin{equation}
p^0=\pm \vert{\bf p}\vert,\quad p^0 =0.
\end{equation}
These ``dispersion relations'' follow from the condition
$\mbox{Det.}\,\,\left({\bf J}\cdot{\bf p} \pm  p^0\right)=0$.

This  situation
immediately raises two problems: ($i$) there are negative energy
solutions, and ($ii$) the equations support solutions with zero energy.
One may either declare that the negative energy solutions, and solutions with
identically vanishing energy content, are to be discarded. Or, face
the possibility that the usual ``quadratic in $\bf E$ and $\bf H$''
expression for the energy density of the electromagnetic field is not
complete.  Here, I argue that the latter is the case by providing an
explicit construct for such an indicated modified expression for the
energy density.

Let us recall a generally accepted way to obtain  the energy density of
the electromagnetic field in vacuum [\refcite{ref5}].

In order to obtain the energy density of the electromagnetic field and the
density of the flux of the electromagnetic energy Landau (see \S 34, p.76
in [\refcite{ref5}])
uses two of Maxwell's equations:
\begin{equation}
\nabla\times{\bf
H}=\frac{4\pi}{c}{\bf j}+\frac{1}{c}\frac{\partial {\bf E}}{\partial t}
\end{equation}
and
\begin{equation}
\nabla\times{\bf E}=-\frac{1}{c}\frac{\partial {\bf H}}{\partial t}
\end{equation}
Landau multiplies both sides of (5) by {\bf E} and both sides of (6) by
{\bf H} and combines the resultant equations:
\begin{equation}
\frac{1}{c}{\bf E}\cdot\frac{\partial{\bf E}}{\partial t}+
\frac{1}{c}{\bf H}\cdot\frac{\partial{\bf H}}{\partial t}=
-\frac{4\pi}{c}{\bf j}\cdot{\bf E}-
({\bf H}\cdot[\nabla\times{\bf E}]-{\bf E}\cdot[\nabla\times{\bf H}])
\end{equation}
Then, using the well-known formula of vectorial analysis, one obtains:
\begin{equation}
\frac{\partial}{\partial t}\left(\frac{E^2+H^2}{8\pi}\right)=
-{\bf j}\cdot{\bf E}-\nabla\cdot{\bf S},
\end{equation}
where the vector
\begin{equation}
{\bf S}=\frac{c}{4\pi}[{\bf E}\times{\bf H}]
\end{equation}
is called the {\it Poynting vector}.

Then Landau integrates (8) over a volume and applies Gauss' theorem to the
second term on {\it rhs}:
\begin{equation}
\frac{\partial}{\partial t}\int\frac{E^2+H^2}{8\pi}d{\cal V}=
-\int{\bf j}\cdot{\bf E}d{\cal V}-\oint{\bf S}\cdot d{\bf f}.
\end{equation}

If the integral, Landau writes further, extends over {\it all} space, then
the surface integral vanishes (the field is zero at infinity). Then one
can express the integral $\int{\bf j}\cdot{\bf E}d{\cal V}$ as a sum $\sum
q{\bf v}\cdot{\bf E}$ over all the charges, and substitute from Eq.(17.7,
[\refcite{ref5}]):  $$ q{\bf v}\cdot{\bf E}=\frac{d}{dt}{\cal E}_{\rm
kin}.
$$
As a result Landau obtains:  \begin{equation}
\frac{d}{dt}\left\{\int\frac{E^2+H^2}{8\pi}d{\cal V}+\sum{\cal E}_{\rm
kin}\right\}=0.
\end{equation}

Thus, Landau concludes, for the closed system consisting of the
electromagnetic field and particles present in it, the quantity in brackets
in this equation is conserved. The second term in this expression is the
kinetic energy (including the rest energy of all particles, of course),
the first term is {\it consequently} the energy of the field {it itself}.
One can therefore call the quantity
\begin{equation}
w=\frac{E^2+H^2}{8\pi}
\end{equation}
the {\it energy density} of the electromagnetic field. Obviously
that {\it it is impossible} to coordinate {\it such} a definition of the
energy density with {\it such}  a configuration of the fields when $w$ is
zero in some point while the fields {\bf E} and {\bf H} are not zero at the
same point.

Here however, we have to make two important comments:

1) Landau uses the transition
$\frac{\partial}{\partial t}\int(...)\rightarrow\frac{d}{dt}\int(...)$ for
a field too freely, without any clarification of this mathematical
operation.

2) Landau states (see [\refcite{ref5}], \S 31 ) that the surface integral $\oint{\bf
S}\cdot d{\bf f}$ vanishes at infinity because the field is zero at
infinity. But in this case one implicitly neglects a radiation field which
can go off to infinity. In  other words, one cannot do the transition
from (10) to (11) without imposing  certain additional conditions, which
prevent this radiation field from going off to infinity. To be more
 specific, let us turn to Landau ([\refcite{ref5}], \S 34, first
footnote): ``{\it Here we also assume that the electromagnetic field of the
 system also vanishes at infinity. This means that, if there is a
radiation of electromagnetic waves by the system, it is assumed that
special `reflecting walls' prevent these waves from going off to
infinity}.''

Let us, now turn to our (and Landau's) formulas (10) and (11):

In classical electrodynamics one assumes that the energy conservation law
is an {\it absolute} law and in order to satisfy this law we must, in
general, take into account a possible {\it change} of energy of these
``reflecting walls'', which {\it may} take place as a result of the
{\it energy exchange} between these ``walls'' and the system ``particles +
fields''.

But we cannot know a mathematically correct way to take into account this
energy in the formula (11) without exact knowledge of the ``nature'' of the
``reflecting walls.'' In this case we cannot obtain an exact energy
conservation law using the concept of the ``reflecting walls.'' In
other words in order to obtain  the {\it exact} energy conservation law
one should not introduce these ``walls,'' but rather we {\it must} assume
that the surface integral $\oint{\bf S}\cdot d{\bf f}$ {\it does not}
vanish at infinity.  But in this case Eq.(10) turns into a trivial
equality, which although satisfying the exact energy conservation law,
cannot be used to  derive any conclusion about
the concrete mathematical form of a energy density of the electromagnetic
field.

$$$$
{\bf 2. Another form of energy density and its connection with  the
Oppenheimer-Ahluwalia zero-energy solutions of the Maxwell equations}

\medskip

\noindent
However,  there is a way to obtain the explicit form of the energy density
of the electromagnetic field. Let me turn to our (with
R.Smirnov-Rueda) papers  [\refcite{ref6},\refcite{ref7}] where we prove
that the electromagnetic field has to be represented by two independent
parts:  \begin{eqnarray} && {\bf E}={\bf E}_0+{\bf E}^*={\bf E}_0({\bf
r}-{\bf r}_q(t))+ {\bf E}^*({\bf r},t),\\ && {\bf H}={\bf H}_0+{\bf
H}^*={\bf H}_0({\bf r}-{\bf r}_q(t))+ {\bf H}^*({\bf r},t).
\end{eqnarray}
Here we note that quasistatic components such as ${\bf E}_0$ and ${\bf
H}_0$ depend only on the distance between the point of observation and the
source position at the instant of observation, whereas time-varying-fields
such as ${\bf E}^*$ and ${\bf H}^*$ depend explicitly on the point of
observation and the time of observation.

Let us now rewrite Eqs. (5) and (6) as formulas (45) and (46) from our
aforementioned paper [\refcite{ref7}]:
\begin{eqnarray} && \nabla\times{\bf H}=\frac{4\pi}{c}{\bf j}+\frac{1}{c}
\frac{d{\bf E}}{dt}\\ && \nabla\times{\bf E}=-\frac{1}{c}\frac{d{\bf
H}}{dt} \end{eqnarray} where the total time derivative of any vector field
value {\bf E} (or {\bf H}) can be calculated by the following rule:
\begin{equation}
\frac{d{\bf E}}{dt}=\frac{\partial{\bf E}^*}{\partial
t}-\Vi{\bf E}_0,
\end{equation}
here ${\bf V}_i$ are velocities of the particles at the same
instant of observation.\fnm{b}\fnt{b}{Note (see [\refcite{ref6},
\refcite{ref7}]) that unlike the
fields $\{\}^*$ the fields $\{\}_0$ {\it do not retard.}}

The mutual independence of the fields $\{\}_0$ and $\{\}^*$ allows us also
to rewrite Eqs. (15) and (16) (taking into account relations (13), (14) and
(17)) as two uncoupled pairs of differential equations:
\begin{eqnarray} && \nabla\times{\bf
H}^*=\frac{1}{c}\frac{\partial{\bf
E}^*}{\partial t},\\ && \nabla\times{\bf
E}^*=-\frac{1}{c}\frac{\partial{\bf B}^*}{\partial t}
\end{eqnarray}
and
\begin{eqnarray} && \nabla\times{\bf
H}_0=\frac{4\pi}{c}{\bf j} -\frac{1}{c}\Vi{\bf E}_0,\\ && \nabla\times{\bf
E}_0=\frac{1}{c}\Vi{\bf H}_0.
\end{eqnarray}

Let us, at last, repeat the calculation of Landau (see above), but now,
taking into account Eqs. (15) and (16) and without imposing
the ``reflecting walls'' type condition.

Let us multiply both sides of (15) by {\bf E} and both sides of (16) by
{\bf H} and combine the resultant equations. Then we get:
\begin{equation}
\frac{1}{c}{\bf E}\cdot\frac{d{\bf E}}{dt}+
\frac{1}{c}{\bf H}\cdot\frac{d{\bf H}}{dt}=
-\frac{4\pi}{c}{\bf j}\cdot{\bf E}-
({\bf H}\cdot[\nabla\times{\bf E}]-{\bf E}\cdot[\nabla\times{\bf H}]).
\end{equation}
Using the rule (17) and the well-known formula of vector analysis, we
obtain:  \begin{eqnarray} \frac{1}{c}{\bf E}\cdot\left\{\frac{\partial{\bf
E}^*}{\partial t}-\Vi{\bf E}_0\right\} & + & \frac{1}{c}{\bf
H}\cdot\left\{\frac{\partial{\bf H}^*}{\partial t}-\Vi{\bf H}_0\right\} =
\nonumber\\
& &
=-\frac{4\pi}{c}{\bf j}\cdot{\bf E}-\nabla\cdot[{\bf
E}\times{\bf H}].
\end{eqnarray}
Then, taking into account the relations (13), (14), and also that
$$
\frac{d\{\}^*}{dt}=\frac{\partial\{\}^*}{\partial t}\qquad
{\rm and}\qquad
\frac{d\{\}_0}{dt}=-\Vi\{\}_0,
$$
and finally, after some transformations we obtain:
\begin{eqnarray}
\frac{\partial}{\partial t}\left(\frac{E^{*2}+H^{*2}}{8\pi}\right) & + &
\frac{d}{dt}\left(\frac{2{\bf E}^*\cdot{\bf E}_0+2{\bf H}^*\cdot{\bf H}_0+
E^2_0+H^2_0}{8\pi}\right)=
\nonumber\\
& &
=-{\bf j}\cdot{\bf E}-\nabla\cdot\left(\frac{c}{4\pi}[{\bf
E}\times{\bf H}]\right).
\end{eqnarray}
Now we can integrate this expression over a volume (taking into account
the relation $q{\bf v}\cdot{\bf E}=\frac{d}{dt}{\cal E}_{\rm kin}$):
\begin{eqnarray}
\frac{\partial}{\partial t}\int\frac{E^{*2}+H^{*2}}{8\pi}d{\cal V} & + &
\frac{d}{dt}\left(\int\frac{2{\bf E}^*\cdot{\bf E}_0+2{\bf
H}^*\cdot{\bf H}_0+ E^2_0+H^2_0}{8\pi}d{\cal V}+\sum{\cal E}_{\rm
kin}\right)= \nonumber\\ & & =-\int\nabla\cdot\left(\frac{c}{4\pi}[{\bf
E}\times{\bf H}]\right)d{\cal V}.
\end{eqnarray}
Let us now extend these integrals over all space and apply Gauss' theorem
to {\it rhs} of (25). In this case, taking into account that the fields
$\{\}_0$ connected with particles vanish at infinity, we obtain:
\begin{equation}
-\int\nabla\cdot\left(\frac{c}{4\pi}[{\bf
E}\times{\bf H}]\right)d{\cal V}\rightarrow
-\oint\left(\frac{c}{4\pi}[{\bf
E}^*\times{\bf H}^*]\right)\cdot d{\bf f}=
-\int\nabla\cdot\left(\frac{c}{4\pi}[{\bf
E}^*\times{\bf H}^*]\right)d{\cal V}.
\end{equation}
It is easy to verify, taking into account Eqs. (18) and (19), that the
last integral in (26) and the first integral in (25) are equal to each
other. Then (25) becomes:
\begin{equation}
\frac{d}{dt}\left(\int\frac{2{\bf E}^*\cdot{\bf E}_0+2{\bf
H}^*\cdot{\bf H}_0+ E^2_0+H^2_0}{8\pi}d{\cal V}+\sum{\cal E}_{\rm
kin}\right)=0.
\end{equation}

We can therefore call the quantity
\begin{equation}
w=\frac{2{\bf E}^*\cdot{\bf E}_0+2{\bf
H}^*\cdot{\bf H}_0+ E^2_0+H^2_0}{8\pi}
\end{equation}
the {\it energy density} of the electromagnetic field.

Note again that one {\it never} can obtain the so called ``{\it
Oppenheimer-Ahluwalia zero energy solutions}'' from  the {\it generally
accepted} form of the electromagnetic energy \begin{equation}
w=\frac{E^2+H^2}{8\pi} \end{equation} because for real fields this
quantity is {\it always} positive and only  can be zero if the fields
{\bf E} and {\bf H} are zero {\it simultaneously}.

But from our {\it new} representation of the density of this energy
\begin{equation}
w=\frac{2
{\bf E}^*
\cdot
{\bf E}_0+2
{\bf H}^*
\cdot{\bf H}_0+ E^2_0+H^2_0}{8}
\end{equation}
it easy to see that the fields $\{\}^*$ and $\{\}_0$ can have  {\it
mutually different} signs because these fields $\{\}^*$ and $\{\}_0$  are
{\it different} fields.  It means that we can have the following relation:
\begin{equation}
2{\bf E}^*\cdot{\bf E}_0+2{\bf
H}^*\cdot{\bf H}_0<0
\end{equation}
and, in turn, we can have a configuration of non-zero fields
for which $w$  is zero:
\begin{equation}
2{\bf E}^*\cdot{\bf E}_0+2{\bf
H}^*\cdot{\bf H}_0= -(E^2_0+H^2_0)
\end{equation}
Actually, it is sufficient that the fields $\{\}^*$ and $\{\}_0$ satisfy
the equations:
\begin{equation}
|{\bf E}^*|=\frac{-|{\bf
E}_0|}{2\cos\alpha} \qquad{\rm and}\qquad |{\bf H}^*|=\frac{-|{\bf
H}_0|}{2\cos\alpha},
\end{equation}
where $\alpha$ is an angle between the
vectors $\{\}^*$ and $\{\}_0$ with the following limits:
\begin{equation}
\frac{\pi}{2}<\alpha<\pi+\frac{\pi}{2}.
\end{equation}
From the formulas (30) and (31) one also can see that there are negative
energy solutions (compare with the remark ($i$) after Eq.(4)).

\clearpage

$$$$
{\bf 3. Conclusion}

\medskip

\noindent
In this short note I do not deal with such concepts as the momentum and
the angular momentum of the electromagnetic field. And although in this
work I use the concept of the Poynting vector, I do not use the
concept of the density of momentum of the field. Let me clarify my point
of view:

On the one hand,  from {\it generally accepted} classical
electrodynamics we know that the Poynting vector is {\it proportional} to
the density of the electromagnetic field momentum. But on the other hand,
paradoxes connected with the Poynting vector exist and they are
well-known.  For example, in our paper [\refcite{ref6}]: if a charge $Q$ is
vibrating in some mechanical way along the $X$-axis, then the value of $w$
(which is a point function like $|{\bf E}|$) on the same axis will be also
oscillating.  The question arises:  {\it how does the test charge $q$
at the point of observation, lying at some fixed distance from the charge
$Q$ along the continuation of the X-axis, ``know'' about the charge $Q$
vibration}?  In other words we have a rather strange situation: the
Poynting vector ${\bf S}=\frac{c}{4\pi}[{\bf E}\times{\bf H}]$ is zero
along this axis (because {\bf H} is zero along this line) but the energy
and the momentum, obviously ``pass'' from point to point along this axis.
This means that we cannot be sure whether using the {\it new} definition of
the energy density will permit use of the {\it old} definition of the
momentum density. This problem, I think, requires very careful
research.  Other quantities of  classical electrodynamics such as
electromagnetic field tensor, electromagnetic energy-momentum tensor etc.
can (and perhaps must) also change their physical meanings.
In fact, a considerable number of works have recently been published which
directly declare: classical electrodynamics must be very sufficiently {\it
reconsidered} \fnm{c}\fnt{c}{See a brilliant review of these works ``Essay
on Non-Maxwellian Theories of Electromagnetism'' by V.V.Dvoeglazov
[\refcite{ref4}].}. To be more specific, let me end this paper with the
words of R.Feynman  who  wrote  [\refcite{ref8}]:  ``...this tremendous
edifice (classical electrodynamics), which is such a beautiful success in
explaining so many phenomena, ultimately falls on its face.  When you
follow any of our physics too far,  you find  that it always gets into
some kind of trouble.  ...the failure  of the  classical electromagnetic
theory.  ...Classical    mechanics   is    a mathematically  consistent
theory; it just  doesn't agree  with experience. It is interesting,
though, that the classical  theory of electromagnetism is an
unsatisfactory theory all by  itself.  There are  difficulties associated
with the  ideas of Maxwell's theory which are not solved by and not
directly associated with quantum mechanics...''

\nonumsection{Acknowledgments}
I am grateful to Dr. D.V.Ahluwalia  for many stimulating discussions and
critical comments.  I acknowledge the brilliant review by Professor
V.Dvoeglazov, which put an idea into me to make the present work. I am
grateful to Zacatecas University for a professorship.

\nonumsection{References}

\end{document}